\documentstyle[12pt,amssymb,amsfonts]{article}

\def \square {\hbox{$\sqcup\!\!\!\!\sqcap$}} 
\newcommand{\be}{\begin{equation}}
\newcommand{\ee}{\end{equation}} 
\newcommand{\bea}{\begin{eqnarray}}
\newcommand{\eea}{\end{eqnarray}}

\begin{document}

\begin{titlepage}

\begin{flushright} 
{\tt 	FTUV/97-74\\ 
	IFIC/97-106\\ }
\end{flushright}

\bigskip

\begin{center}

{\bf{\LARGE General Covariance and Free Fields\\

in Two Dimensions.}}
\footnote{Work partially supported by the 
{\it Comisi\'on Interministerial de Ciencia y Tecnolog\'{\i}a}\/ 
and {\it DGICYT}}

\bigskip 

 J.~Cruz \footnote{\sc cruz@lie.uv.es},
 D. J. Navarro \footnote{\sc dnavarro@lie.uv.es},
and J.~Navarro-Salas \footnote{\sc jnavarro@lie.uv.es}.

\end{center}

\begin{center}

\footnotesize
	 Departamento de F\'{\i}sica Te\'orica and 
	IFIC, Centro Mixto Universidad de Valencia-CSIC.
	Facultad de F\'{\i}sica, Universidad de Valencia,	
        Burjassot-46100, Valencia, Spain. 
\end{center}

\normalsize 

\bigskip
\bigskip


\begin{center}
			{\bf Abstract}
\end{center}
We investigate the canonical equivalence of a matter-coupled 2D dilaton gravity
theory defined by the action functional $S=\int d^2x \sqrt{-g}
 \left( R\phi +
V \left( \phi \right)\right.$
$\left.- {1\over2} H \left( \phi \right) \left( \nabla
f\right)^2\right)$, and a free field theory.
When the scalar field $f$ is minimally coupled to the metric field
$(H(\phi)=1)$ the theory is equivalent, up to a boundary contribution,
to a theory
of three free scalar fields with indefinite kinetic terms, irrespective of the
 particular form of the potential $V\left(\phi\right)$. 
 If the potential is an exponential
function of the dilaton one recovers a generalized form of the classical
 canonical transformation of Liouville theory.
When  $f$ is a dilaton coupled scalar $(H(\phi)=\phi)$ and the potential is an
arbitrary power of the dilaton the theory is also canonically equivalent to a
theory of three free fields with a Minkowskian target space.
In the simplest case $(V(\phi)=0)$ we provide an explicit free field
realization of the Einstein-Rosen midisuperspace. The Virasoro
anomaly and the consistence of the Dirac operator quantization play a central
role in our approach.\\

\noindent PACS number(s): 04.60.Kz, 04.60.Ds

\end{titlepage}
 
\newpage

\section{Introduction}
One of the main features of the Liouville field theory is that a canonical 
transformation maps the theory into a free field theory \cite{Braaten,D'Hoker,
Gervais}. The free field realization of the interacting theory allows
to carry out an operator quantization of the model although a complete
understanding of the full quantum theory remains elusive \cite{Kazama}.
The conformal invariance of the theory can be understood, in a natural way,
in terms of a generally covariant action \cite{Jackiw} which provides an 
improved, traceless energy momentum tensor. 
In conformal gauge the generally covariant action describes a Liouville field 
 together with a free field which, due to general
  covariance, possesses the same improvement term
as the Liouville field but with opposite contribution to the hamiltonian constraint. 
 At the quantum level the hamiltonian 
and momentum constraints generate two copies of the Virasoro algebra but
there exists an ambiguity to determine the value of the central charge
as it has been pointed out recently \cite{Cangemi} in the context of the CGHS
theory \cite{CGHS}. The string-type
 quantization is insensitive to the signature  
 of the target space and produces a non-vanishing central charge. However, 
 in the functional Schr\"odinger quantization, the 
 total value of the Virasoro anomaly vanishes and one can consistently impose all
  the Virasoro quantum constraints \cite{Cangemi}.  Remarkably, it has been possible to explicitly solve
  all the Virasoro quantum constraints for a generic model of two-dimensional dilaton
  gravity in the functional Schr\"odinger representation
\cite{Louis}. Turning things round one
could expect that the possibility of finding solutions to the quantum
constraints could reflect the fact of an anomaly cancellation between two free
fields.
Therefore one could guess that the above discussion on the
Liouville theory could be extended to a generic
  theory of 2D dilaton gravity. In fact the canonical transformation mapping the CGHS
  model into a free field theory, both on-shell \cite{Verlinde} and off-shell 
  \cite{Cangemi,kuc}, can be seen as a limiting
  case of a canonical transformation converting the generally covariant 
  (Liouville) theory
  into a free field theory with a Minkowskian target space.
  The aim of this paper is to investigate the validity of this conjecture for
  a large class of
  generally covariant theories
  involving a two-dimensional metric, a dilaton and a
  single massless scalar field.\\
  
  In section 2 we shall analyze the covariant 
  theory associated with the Liouville action showing how it can be transformed 
  into a theory of two free fields without any improvement term. This way of presenting 
  the canonical structure of the theory suggests that the
   underlying reason to transform
  it into a free field theory is general covariance, thus supporting the motivation
   of this work.
   In section 3 we shall demonstrate,
   completing the construction sketched in Ref. \cite{Cruz}, that a first
    order generic theory of 2D 
   dilaton gravity minimally coupled to a massless scalar field can be mapped into
   a theory of free fields with a Minkowskian target space.
   We shall also address the problem of the comparison of the 
   Schr\"odinger quantization
   in geometrical variables and in terms of free field variables.
  It is well-known that the quantization procedure does 
  not commute with the canonical transformations and it has to be 
  checked on a case-by-case basis whether the quantum wave functionals
   obtained with one set of variables are 
   related by unitarity with the quantum wave functionals
   derived with another set of canonical variables.
   This question has been analyzed for the 
   matter free CGHS model in \cite{Benedict, Louis2} with a positive answer
   and we shall prove that for a generic pure gravity model
    the quantum wave functionals in geometrical variables are equivalent to those obtained
    with the free field variables.
This result reenforces the idea that the free field representation
provides an adequate quantization of two-dimensional dilaton gravity
theories.\\
   
In the remaining part of this paper we shall investigate whether the result of
sections 2 and 3 can be extended to theories of 2D dilaton gravity with a
non-minimal coupling to a matter field.
   In section 4 we study the simplest situation:
   the CGHS model coupled to a Liouville field.
    Using previous results it is clear that
   the dilaton-gravity sector is equivalent to two free fields and the remaining
   Liouville-gravity sector is also equivalent to two free fields.
   However it is not clear that the full theory can be transformed into a theory with
  three free fields. 
  We shall show that this is the  case and, in contrast with minimally
  coupled matter field models, it is necessary to mix in a non-trivial
  way the metric-dilaton and matter fields to produce a theory with 
  three free fields.
   In section 5 
   we consider a more realistic family of models with a dilaton coupled scalar.
   This way one consider important cases of dimensionally reduced 
   general relativity, which has been recently studied from a path integral approach 
   with different results \cite{Bousso}.
   We shall start our study analizing with detail the midisuperspace model of
   cylindrically symmetric gravitational waves with one polarization
   \cite{Einstein}. The canonical analysis of this system was initiated in 
   Ref. \cite{Kuchar0}
   and it is mostly studied in the framework of the reduced phase-space
   quantization \cite{Ashtekar}. In this approach, first solve the constraints
   and then quantize, the possible anomalies
   of the constraint (Virasoro) algebra do not arise and it could imply the
   scheme be inequivalent to the standard Dirac quantization. We shall prove
   that the theory
   can also be mapped into a theory of three free fields with a Minkowskian target space,
   thus generalizing the results of the reduced phase-space
   approach. Moreover we shall show that the free field
   realization of the Einstein-Rosen midisuperspace can be extended to models 
   with a potential of the form $V(\phi) \propto \phi^a$, where $a$ is an arbitrary
   real parameter. This includes the important case of spherically reduced
   Einstein gravity. Finally, in section 6 we shall state our conclusions.

\section{Canonical Structure of the generally covariant Liouville Action}
The Liouville action can be obtained from the generally covariant action
\be
\int d^2x \sqrt{-g} \left( \beta \left(\nabla \phi\right)^2 + 4\lambda^2
e^{2\beta\phi} + R \phi \right) \>.
\label{ai}
\ee
The equation of motion $R=0$ allow us to fix the gauge $\rho=0$ in conformal
coordinates ($ds^2 = -e^{2\rho} dx^{+} dx^{-}$). The equation of motion of
the field $\varphi = 2 \beta \phi$ is then the Liouville equation
\be
\partial_{+} \partial_{-} \varphi + 2 \lambda^2 \beta e^{\varphi} = 0 \, .
\label{aii}
\ee
To study the generally covariant theory it is convenient to parametrize the
metric in the form
\be
ds^2 = e^{2\rho} \left[ -u^2dt^2 + (vdt+dx)^2 \right] \, ,
\label{aiii}
\ee
where the functions $u$ and $v$ are related to the shift and lapse functions.
In terms of the fields $\varphi = 2\rho + 2\beta\phi$ and $\eta = 2\rho $ the
hamiltonian form of the action is
\be
S = \int d^2x (\pi_{\eta} \dot{\eta} + \pi_{\varphi} \dot{\varphi} - uH
-vP) \, ,
\label{aiv}
\ee
where the hamiltonian and momentum constraints read as
\bea
H &=& -(\beta\pi_{\varphi}^2 + \frac{1}{4\beta}\varphi'^2 - 4\lambda^2
e\sp\varphi - \frac{1}{\beta}\varphi'') + \beta\pi_\eta\sp2 + 
\frac{1}{4\beta}\eta'\sp 2-\frac{1}{\beta}\eta'' \, , \label{av} \\
P &=& \pi_\varphi \varphi'+\pi_\eta\eta'-2\pi_\varphi'-2\pi_\eta' \, .
\label{avi}
\eea
It is well-known that one map the Liouville field $\varphi$ into a free field
$\psi$ by exploiting the form of the classical solution of the Liouville
equation (\ref{aii})
\be
\varphi = \log \frac{\partial_{+} A_{+}(x^{+}) \partial_{-} A_{-}(x^{-})}
{(1+\lambda^2 \beta A_+ A_-)^2} \, ,
\label{avii}
\ee
where $A_+$ and $A_-$ are two arbitrary functions. The free field $\psi$ can be
constructed as a linear combination of chiral functions but the explicit form
is not unambiguous. We choose a definition closely related to the one given in
\cite{D'Hoker}
\bea
\varphi &=& \psi - 2\log (1+\lambda\sp 2\beta A_+ A_-) \, , \label{aviii}\\
\pi_\varphi& = & \pi_\psi - \lambda\sp 2\frac{A_+'A_--A_+A_-'}{(1+\lambda\sp 
2\beta A_+A_-)} \, ,
\label{aix}
\eea
with
\be
(\log A'_{\pm})' = \frac{\psi'}{2} \pm \beta \pi_\psi \, .
\label{ax}
\ee
The above expressions define a canonical transformation and a straightforward
way to check this is to compute the canonical 2-form of the Liouville theory
in terms of $A_+$ and $A_-$
\be
\omega_{L} = \int dx\ \delta\varphi \wedge \delta\pi_{\varphi} \, ,
\label{axi}
\ee
where $\delta$ stands for the exterior derivative on phase-space.
 After same computation one obtains (from now on
we shall omit the exterior product)
\be
\omega_{L} = \frac{1}{\beta}\int \delta(\log A'_+ A'_-)\delta(\log\frac{A'_+}{A'_-})' +
\omega_{b} \, ,
\label{axii}
\ee
where $\omega_b$ is a total derivative term\footnote{This term vanishes when
the functions $A_{\pm}$ have a diagonal monodromy.}
\be
\omega_b = -2\lambda^2 \beta \int d \left[ \frac{\delta (A_+A_-)
\delta (\log \frac{A'_+}{A'_-})+ \delta A_{+} \delta A_{-}}{1+\lambda^2 \beta 
A_+A_-} \right] \, .
\label{axiii}
\ee
In terms of $\psi$ and $\pi_{\psi}$, and up to a total derivative, the
symplectic 2-form of the theory adopts the canonical form
\be
\omega_L = \int dx \ \delta \psi \delta \pi_{\psi} \, .
\label{axiv}
\ee
Going back to the generally covariant theory and, in terms of the canonical
variables ($\psi$, $\pi_\psi$; $\eta$, $\pi_\eta$), the constraints become
\bea
H&=&-(\beta\pi_\psi\sp 2+\frac{1}{4\beta}\psi'\sp 
2-\frac{1}{\beta}\psi'')+\beta\pi_\eta\sp 2+\frac{1}{4\beta}\eta'\sp 
2-\frac{1}{\beta}\eta''\quad ,\label{axv}\\
P&=&\pi_\psi \psi' + \pi_\eta\eta' -2\pi_\psi'-2\pi_\eta'\quad .
\label{axvi}
\eea
Therefore, the theory is described by two improved free fields with opposite
contributions to the hamiltonian constraint. The gauge fixing $\eta=0=\pi_\eta$
recovers the standard results but due to the presence of the field $\eta$ one can
simplify further the constraints using new canonical transformations. First
we shall introduce the canonical variables ($\bar{X}^{\pm}$, $\bar{\Pi}_{\pm}$)
\cite{Kuchar} defined as
\bea
2\bar{\Pi}_{\pm} &=& -   \sqrt{2\beta} (\pi_\eta + \pi_\psi) \mp \frac{1}
{\sqrt{2\beta}} (\eta' - \psi') \, ,
\label{axvii}         \\
2\bar{X}^{\pm\prime}  &=& \mp \sqrt{2\beta} (\pi_\eta - \pi_\psi) - \frac{1}
{\sqrt{2\beta}} (\eta' + \psi') \, ,
\label{axviii}
\eea
in terms of which the constraints $C_{\pm} = \pm{1\over2} (H\pm P)$ become
\be
C_{\pm} = \bar{X}^{\pm\prime}\bar{\Pi}_{\pm} + \sqrt{\frac{2}{\beta}}
\bar \Pi^{\prime}_{\pm} \, ,
\label{axix}
\ee
and a further canonical transformation similar to the one introduced in \cite{Cangemi}
\bea
\bar{X}^{\pm} &=& -\sqrt{\frac{2}{\beta}} \log X^{\pm\prime} \, , \label{axx}\\
\Pi_{\pm} &=& -\sqrt{\frac{2}{\beta}} \frac{X^{\pm\prime}}{(X^{\pm\prime\prime})^2}
\bar{\Pi}_{\pm} + \sqrt{\frac{2}{\beta}} \frac{\bar \Pi^{\prime}_{\pm}}{X^{\pm\prime}} \, ,
\label{axxi}
\eea
removes the ``improvement'' terms of the energy-momentum tensor
(\ref{av}),(\ref{avi})
\be
C_{\pm} = X^{\pm\prime} \Pi_{\pm} \, .
\label{axxii}
\ee
Finally, the transformation 
\bea
2\Pi_{\pm} &=& - (\pi_0 + \pi_1) \mp (r^{0\prime} - r^{1\prime}) \, ,
\label{axxiii}\\
2X^{\pm\prime}  &=& \mp (\pi_0 - \pi_1) - (r^{0\prime} + r^{1\prime}) \, ,
\label{axxiv}
\eea
brings the constraints into those of two scalar free fields of opposite
signature
\bea
H &=& \frac{1}{2}\left(\pi_0^2 + \left(r^{0\prime} \right)^2\right)
 - \frac{1}{2}\left(\pi_1^2 + \left(r^{1\prime}\right)^2\right) \, ,
 \label{axxv}\\
P &=& \pi_0 r^{0\prime} + \pi_1 r^{1\prime} \, .
\label{axxvi}
\eea
In the standard quantization of conformal field theory each scalar field
contributes with $c=1$ to the total value of the Virasoro anomaly. Therefore,
the constraints have a non-trivial center and the physical states cannot be
annihilated by all the Virasoro quantum operators.
A proper quantization requires the introduction of ghost fields and background
charges to achieve a total zero center \cite{Polyakov}.
However, explicit solutions to the quantum constraints were constructed
in Ref. \cite{Louis} and the way out to the apparent contradiction was provided, in the 
context of the CGHS theory, in Ref. \cite{Cangemi}.
In the functional
Schr\"odinger representation, where the states have 
manifestly positive norms, the scalar
with negative kinetic energy has a opposite definition of creation and
annihilation operators and contributes with $c=-1$ to the Virasoro anomaly.
So, the contributions of the two scalar fields cancel
and one can consistently impose all the
Virasoro constraints.
A similar result can be obtained by evaluating the contribution to the
 central charge of the fields $\varphi$ 
and $\eta$. Due to the presence of "background charges" in (\ref{axv}),(\ref{axiv})
we have $c_{\varphi}=1+3Q^2$ and $c_{\eta}=-1-3Q^2$, in the Schr\"odinger
representation while $c_{\eta}=1-3Q^2$ in the standard conformal field 
quantization.
In the Schr\"odinger representation one gets a zero center $c=c_{\varphi}+
c_{\eta}=0$, while one obtains $c=2$ in the standard quantization approach.
In general, the BRST and the functional Schr\"odinger quantization give inequivalent physical
spectrum \cite{Benedict2}. \\

The fact that the solutions for the quantum constraints given in \cite{Louis}
are valid for a generic theory of pure dilaton-gravity suggests that the above
mechanism for getting a vanishing central charge could also work for a generic
theory. This question will be considered in the next section.

\section{Canonical Structure of a Generic Model of 2D dilaton gravity 
minimally coupled to a scalar field}

In the previous section we have seen that the generally covariant Liouville
 theory can be converted,
 by means of a canonical transformation, into a free field theory
 with a target space of indefinite signature. We have obtained this 
 result by composing the classical canonical transformation
 of Liouville theory with some additional canonical transformations.
 One can also get this result in a different way by using the form
  of the classical solutions
  of the covariant theory expressed in terms of four arbitrary chiral functions
 \cite{Cruz2}. This way the generally covariant Liouville theory
  and the Jackiw-Teitelboim model \cite{jack}, which can be regarded
  as particular cases of a large family of 2D dilaton gravity models
  \cite{Banks} can be explicitly transformed
 into a theory of free fields. 
 It is well-known that by a conformal reparametrization of the fields, one
  can get rid of the kinetic term of the dilaton and bring the action of
  a
 generic model of 2D dilaton gravity into the form
 \be
 S=\int d^2x\sqrt{-g}\left(R\phi+\lambda^2V\left(\phi\right)-{1\over2}\left(\nabla
 f\right)^2\right) 
 \>,\label{bi}
 \ee
 where we have added, for convenience, a minimally coupled massless 
 scalar field $f$, $\lambda^2$ is a dimensionfull parameter and $V(\phi)$
 is a dimensionless arbitrary function of the dilaton. For the
  exponential (Liouville) model we have
 $V(\phi)=4e^{\beta\phi}$, while $V(\phi)=4$ for the CGHS
  model and $V(\phi)=4\phi$
 for the Jackiw-Teitelboim model. The spherically reduced
  Einstein gravity can also
 be seen as a 2D dilaton gravity model with $V(\phi)={2\over\sqrt{\phi}}$.
 Restricting the 4D metric to be
 spherically symmetric
 \be
 ds^2_{(4)}=g_{\mu\nu}dx^{\mu}dx^{\nu}+{\phi^2\over\lambda^2}d\Omega^2
 \>,
 \ee
 where $x^{\mu}$ are coordinates on a two-dimensional spacetime with 
 metric $g_{\mu\nu}$, ${\phi\over\lambda}$ is the radial coordinate
 and $d\Omega^2$ is the line element of the 2-sphere
with area $4\pi$, the dimensionally reduced Hilbert-Einstein action 
 is
 \be
 S=\int d^2x \sqrt{-g}\left(R\phi+2\left(\nabla\phi\right)^2+2\lambda^2\right)
 \>,
 \ee
 and a conformal reparametrization of the 2D metric leads to an action
 of the form
 \be
 S=\int d^2x \sqrt{-g}\left(R\phi+{2\lambda^2\over\sqrt\phi}\right)
 \>.
 \ee

 It is interesting to point out now that, for the Liouville model, the bulk part of the symplectic 
 two-form (\ref{axii}) and the constraints (\ref{axv}),(\ref{axvi}) are independent
 of the coupling parameter $\lambda^2$.
 It only appears in the boundary term $\omega_b$ of the symplectic form.
 This also happens for the Jackiw-Teitelboim model \cite{Cruz2} and suggests that the free 
 field behaviour of $\rho$ and $\phi$ in the theory
 \be
 S=\int d^2x\sqrt{-g}R\phi
 \>,
 \ee
 could be transplanted to the fields
 $\rho\left(x^{\pm},\lambda^2=0\right)$ and $\phi
 \left(x^{\pm},\lambda^2=0\right)$ for a generic model and that the exact expressions of
 $\rho$ and $\phi$, in terms of chiral functions, could define a proper
 canonical transformation.
 To elaborate this idea we shall start our analysis parametrizing 
 the two-dimensional metric as in 
 (\ref{aiii}). The canonical form of the action (\ref{bi}) then reads as
 \be
  S=\int d^2x\left(\pi_{\rho}\dot \rho+\pi_{\phi}\dot\phi+\pi_f\dot f
  -uH-vP\right)
  \>,\label{bii}
  \ee
 where 
  \be
   H=-{1\over2}\pi_{\rho}\pi_{\phi}+2\left(\phi^{\prime\prime}-\phi^{\prime}\rho^{\prime}
 \right)-e^{2\rho}\lambda^2V\left(\phi\right)+{1\over2} 
 \left(\pi_{f}^2+f^{\prime 2}\right)\>,\label{biii}
 \ee
  \be
 P=\rho^{\prime}\pi_{\rho}-\pi_{\rho}^{\prime}
 +\phi^{\prime}\pi_{\phi}+\pi_{f}f^{\prime}\>.\label{biv}
 \ee
 Mimicking the idea of Liouville theory we want to use the expression of the 
 classical solutions in terms of chiral functions to promote them into a
 canonical transformation.
 In conformal gauge the system of equations of motion is
 \be
 8e^{-2\rho}\partial_+\partial_-\rho=-\lambda^2V^{\prime}\left(\phi\right)\>,\label{bv}
 \ee
 \be
 -4e^{-2\rho}\partial_+\partial_-\phi=\lambda^2V\left(\phi\right)\>,\label{bvi}
 \ee
 \be
- \partial_{\pm}^2\phi+2\partial_{\pm}\phi\partial_{\pm}
 \rho=T_{\pm\pm}^f={1\over2}\left(\partial_{\pm}f\right)^2
 \>,\label{bvii}
 \ee
  \be
  \partial_+\partial_-f=0
  \>,  \label{bviii}
  \ee
  but the problem is that, up to specific models, the 
  general solution is unknown. To bypass this situation one can first
restrict the 
  theory imposing chirality to the scalar matter field. Assuming that $T^f_{--}=0$ 
  one can identify the
  two independents chiral functions of the nontrivial sector of the theory. 
  It is easy to see that
  equation (\ref{bvii}) implies
 \be
e^{-2\rho}\partial_-\phi=a\>,    \label{bix}
 \ee
where $a$ is an arbitrary function of the $x^+$ coordinate.  
Inserting this equation into (\ref{bvi}) we get
\be
\partial_+\partial_-\phi+{\lambda^2\over4a}V(\phi)\partial_-\phi=0
\>,\label{bx}
\ee
which has the solution
 \be
 \partial_+\phi=A-{\lambda^2\over 4a}J\left(\phi\right)\>.\label{bxi}
  \ee
 with $A$ another arbitrary function of the $x^+$ coordinate
 and $J^{\prime}\left(\phi\right)=V\left(\phi\right)$.    
Equations (\ref{bvii}),(\ref{bix}),(\ref{bxi}) we can evaluate the non-trivial 
component of the energy momentum tensor in terms of the fields $A$ and $a$
 \be
 C_+=2\left(\partial_+A+{A\over a}\partial_+ a\right)
 \>.\label{bxii}
 \ee
For arbitrary $J(\phi)$ it is not possible to integrate 
explicitly equation (\ref{bxi}).
However, the implicit solution for $\phi$ in terms of $A,a$ and a 
constant of integration $\beta$, which is in this case an arbitrary function of the
$x^-$ coordinate, 
contains enough information to work out the symplectic
2-form of the theory
\be
\omega=\int dx \ \delta\phi\delta\pi_{\phi}
+\delta\rho\delta\pi_{\rho}+\delta f\delta \pi_f
\>,\label{bxiii}
\ee
where the canonical momenta are given by
\be
  \pi_{\phi}=-2
 \dot\rho=
  {\lambda^2\over4  a}V\left(\phi\right)+{ a^{\prime}\over  a}
  -\left({\partial^2_-\phi\over\partial_-\phi}\right)
  \>,\label{bxiv}
\ee
\be
  \pi_{\rho}=-2\dot \phi=
  -2A +{\lambda^2\over2  a}J\left(\phi\right)
  -2\partial_-\phi\>.\label{bxv}
  \ee
 Inserting this into (\ref{bxiii}) we get
 \bea
 \omega&=&-\int dx\left[-\delta\phi{\delta a^{\prime}\over a}+{a^{\prime}\over a^2}\delta\phi\delta a
 +\delta\phi{\partial^2_-\delta\phi\over\partial_-\phi}-
 {\partial_-^2\phi\over\left(\partial_-\phi\right)^2}\delta\phi
 \partial_-\delta\phi\right. \nonumber\\
 &&+{\partial_-\delta\phi\over\partial_-\phi}\delta A
 +\lambda^2J\left(\phi\right){\partial_-\delta\phi\over\partial_-\phi}
 {\delta a\over4a^2}-
 {\lambda^2\over4a}V\left(\phi\right){\partial_-\delta\phi\over\partial_-\phi}
 \delta\phi-{\delta a\over a}\delta A\nonumber\\
 &&\left.-{\delta a \over a}\partial_-\delta\phi+\delta f\delta \pi_f\right]
 \>. \label{bxvi}
\eea
Taking into account the following identities
\be
-\delta\phi{\delta a^{\prime}\over a}= -\left(\delta\phi{\delta a\over a}\right)^{\prime}
+\delta A{\delta a\over a}-{\lambda^2\over 4a^2}V\left(\phi\right)\delta\phi\delta a 
-\partial_-\delta\phi{\delta a\over a}
-{a^{\prime}\over a^2}\delta\phi\delta a
\>,\label{bxvii}
\ee
\bea
\delta \phi{\partial_-^2\delta\phi\over\partial_-\phi}&=&-
\left(\delta\phi{\delta\partial_-\phi\over\partial_-\phi}\right)^{\prime}
+\delta\phi{\partial_+\partial_-\delta\phi\over\partial_-\phi}
+{\delta\partial_+\phi\delta\partial_-\phi\over\partial_-\phi}\nonumber\\
&&-{\partial_+\partial_-\phi
\over\left(\partial_-\phi\right)^2} \delta\phi\partial_-\delta\phi
+{\partial^2_-\phi\over\left(\partial_-\phi\right)^2}\delta\phi\partial_-\delta\phi
\>,\label{bxviii}
\eea
which can be checked by using the relation (\ref{bxi}), the
infinite-dimensional part of the symplectic form turns out to be independent
 of the coupling parameter $\lambda^2$ and
can be expressed in terms of the fields $A$, $a$, $f$ and $\pi_f$ only
\be
\omega=\int dx\ 2{\delta a\over a}\delta  A +\delta f\delta \pi_f
+\int d\left(\delta\phi{\delta a\over a}+\delta\phi{\delta\partial_-\phi
\over\partial_-\phi}\right)
\>.\label{bxix}
\ee
The parameter $\lambda^2$ only appears in the boundary term
in an implicit way through
the relation $\phi=\phi\left(A,a;\lambda^2\right)$
defined by the equation (\ref{bxi}).
It is then clear that defining the canonical variables
\be
X^+=\log aA, \qquad \Pi_+=2A
\>,\label{bxx}
\ee
the symplectic form takes, up to a boundary term, the form
\be
\omega=\int dx\ \delta X^+\delta \Pi_++\delta f\delta \pi_f
\>,\label{bxxi}
\ee
and the non-trivial constraint $C_+={1\over2}\left(H+P\right)$ takes the free field form
\be
C_+=\Pi_+X^{+\prime}+{1\over4}\left(\pi_f+f^{\prime}\right)^2
\>.\label{bxxii}
\ee
At this point we must stress that in the above discussion we have made use of the 
unconstrained equations of motion (\ref{bv}),(\ref{bvi}).
To define an off-shell canonical transformation we have to check that the above derivation is still valid
if the functions $A$, $a$, $\beta$ are arbitrary functions of the space time coordinates.
To this end we introduce the following definition.
The symbol  $\ \widetilde{}\ $  affecting any functional of 
 the previous chiral functions $A,a,\beta$ means that 
 they are considered as arbitrary (not chiral) functions
  $\tilde A,\tilde a,\tilde \beta$
  and that the possible derivatives 
 or integrations have been replaced according to the rule
 ${\partial_{\pm}}\longrightarrow \pm \partial_x\ \left(\partial_{\pm}^{-1}
 \longrightarrow \pm\partial_x^{-1}\right)$ as in passing from the solution (\ref{avii})
 of Liouville theory into the canonical transformation (\ref{aviii}),(\ref{aix}).
Therefore, we define now a transformation to the new set of 
variables $\tilde A,
 \tilde a,\tilde \beta $
 \be
  \phi=\tilde\phi\>,\label{bxxiii}
  \ee
  \be
  \pi_{\phi}=-2\widetilde
 { \dot\rho}=
  {1\over4 \tilde a}V\left(\tilde\phi\right)+{\tilde a^{\prime}\over \tilde a}
  -\widetilde{\left({\partial^2_-\phi\over\partial_-\phi}\right)}
  \>,\label{bxxiv}
  \ee
  \be
  \rho={1\over2}\log {\widetilde{\partial_-\phi}\over \tilde a}\>,\label{bxxv}
  \ee
  \be
  \pi_{\rho}=-2\widetilde{\dot \phi}=
  -2\tilde A +{1\over2 \tilde a}J\left(\tilde\phi\right)
  -2\widetilde{\partial_-\phi} \>,\label{bxxvi}
  \ee
where from now on we absorb the $\lambda^2$ parameter in the potential function
$V(\phi)$.
Since the dependence of $\phi$ on $\beta$ is
 ultralocal we can prove immediately the following identities

\be
\left(\tilde\phi\right)^{\prime} =\widetilde{\partial_+\phi}-\widetilde{
\partial_-\phi}\>, \label{bxxvii}
\ee
\be
\left(\widetilde{\partial_-\phi}\right)^{\prime}=\widetilde{\partial_+
\partial_-\phi}-\widetilde{\partial^2_-\phi}\>.\label{bxxviii}
\ee
To prove (\ref{bxxvii}) we expand $\left(\tilde \phi\right)^{\prime}$ in terms of
$\tilde A,\tilde a,\tilde \beta$
\bea
\left(\tilde\phi\right)^{\prime}&=&\sum_i{\partial \tilde \phi\over\partial
 \tilde A^{(i)}}
\left(\tilde A\right)^{(i+1)}+{\partial\tilde\phi\over\partial\tilde \beta}
\left(\tilde \beta\right)^{\prime}=\sum_i
\widetilde{{\partial\phi\over\partial\left(\partial_+^{(i)}A\right)}
\partial_+^{(i+1)}A}-\widetilde{{\partial\phi\over\partial\beta}\partial_- \beta}
\nonumber     \\
&& = \widetilde{\partial_+\phi}-\widetilde{\partial_-\phi}  \>, \nonumber
\eea
where the superindex $\ ^{(i)}$ indicates the order of derivation. 
Equation (\ref{bxxviii}) can be checked in an analogous way.
  These two identities imply that the computation of the constraints and the symplectic form in terms of 
$A,a$ can be extended to the fields $\tilde A,\tilde a$ and therefore all the above results
are valid when $A$ and $a$ are replaced by $\tilde A$ and $\tilde a$.\\

We shall now return to the general situation.
Without assuming chirality the solution to the equations of motion
is parametrized by four arbitrary chiral functions
  $A\left(x^+\right),a\left(x^+\right),
  B\left(x^-\right), 
 b\left(x^-\right)$.
  They have a very simple interpretation. Two of them are the two arbitrary functions 
  associated with the residual conformal coordinate transformations while the other 
  two account for the two components of the energy-momentum tensor.
  It is convenient to choose $A,a,B,b$ in such a way that when $T^f_{--}=0$ 
  the equations
  of motion are equivalent to (\ref{bix}),(\ref{bxi})
   and when $T^f_{++}=0$ they are equivalent to
   \be
 \partial_-\phi=B-{1\over4 b}J\left(\phi\right)\>,\label{bxxix}
 \ee
 \be
 e^{-2\rho}\partial_+\phi=b
 \>.\label{bxxx}
 \ee
  The classical solution
  $\phi=\phi\left(A,a;B,b\right)$,
  $\rho=\rho\left(A,a;B,b\right)$
  can be employed to construct a transformation to the new variables
  $A,a,B,b$
   \be
   \phi=\tilde \phi,\ \pi_{\phi}=-2\widetilde{\dot\rho},
\ \rho=\tilde\rho,\ \pi_{\rho}=-2\widetilde{\dot\phi}\>.
\label{bxxxi}
\ee
 This transformation 
 reduces to (\ref{bxxiii}-\ref{bxxvi}) when $T^f_{--}=0$ and to the analogous
  chiral transformation when $T^f_{++}=0$. 
Due to general covariance the constraints
$C_{\pm}=\pm{1\over2}\left(H\pm P\right)$
are chiral functions on-shell and therefore $C_+\left(C_-\right)$ must take 
the same form as when we impose the condition $C_-=0\left(C_+=0\right).$
This is so because restriction to chirality (e.g. $T^f_{--}=0$) gives
 a condition over the fields
$X^-,\Pi_-$ and keeps free $X^+,\Pi_+$.
Therefore any correction to the constraint $C_+$ should depend only on the fields
$X^-,\Pi_-$ in order to get the adequate chiral limit.
But over solutions the fields $X^{\pm},\Pi_{\pm}$ must depend only on the 
$x^{\pm}$ coordinate respectively so such a correction would not be consistent.
So in accordance with our previous result we have
\be
C_{\pm}=\Pi_{\pm}X^{\pm\prime}\pm{1\over4}\left(\pi_f\pm f^{\prime}\right)^2
 \>,\label{bxxxii}
 \ee
where $X^+,\Pi_+$ are given by (\ref{bxx}) and
\be
X^{-}=\log \tilde b \tilde B,
\qquad \Pi_-=2\tilde B\>. \label{bxxxiii}
\ee
Now the question to be considered is the canonicity
of the transformation
 $\left(\phi,\pi_{\phi},\rho,\pi_{\rho}\right)\longrightarrow\left(X^{\pm},\Pi_{\pm}
 \right)$(up to a boundary term).
 In doing so let us write the most general expression
  for the symplectic 2-form $\omega$
 in terms of $X^{\pm},\Pi_{\pm}$
 \bea
 \omega&=& \int dxdy\left[F\delta X^+(x)\delta\Pi_+(y)+
 G\delta X^-(x)\delta\Pi_-(y)
 +
 L\delta X^+(x)\delta X^-(y)\right.\nonumber\\&&
 \left.+M\delta X^+ (x)\delta\Pi_-(y)
 +N\delta X^-(x)\delta\Pi_+(y)
+ Q\delta\Pi_+(x)\delta\Pi_-(y)\right]
 \>. \label{xxxiv}
 \eea
 In order to reproduce the hamiltonian equations of motion
 in conformal gauge $\partial_{\mp}X^{\pm}=\partial_{\mp}
 \Pi_{\pm}=0$, it is clear that the
 hamiltonian vector field $X_H$ must be
 \bea
 X_H&=&\int dx\left[X^{+\prime}(x){\delta\over\delta X^+(x)}-
 X^{-\prime}(x){\delta\over\delta X^-(x)}
 +\Pi_+^{\prime}(x){\delta\over\delta\Pi_+(x)}\right.\nonumber\\
 &&\left.-\Pi_-^{\prime}(x){\delta\over \delta\Pi_-(x)}\right]
                             \>. \label{bxxxv}
 \eea
 The condition
 \be
 i_{X_H}\omega=-\delta H
 \>,\label{bxxxvi}
\ee
together
with the requirement that the symplectic form be invariant
 under spatial diffeomorphisms (i.e, the integral of a scalar density),
 implies that
 the transformation is canonical, up to a boundary term $\omega_b$
\be
\omega=\int dx \ \delta X^+\delta \Pi_+ +
\delta X^-\delta \Pi_- +\omega_b
\>.
\ee
Therefore the canonical form of the action is of the form
\bea
S&=&\int d^2x\left[\Pi_+\dot X^++\Pi_-\dot X^-+\pi_f\dot f \right.\nonumber\\
&&-u\left(\Pi_+
X^{+\prime}-\Pi_-X^{-\prime}+{1\over2}\left(\pi_f^2+f^{\prime 2}\right)\right)
\nonumber\\
&&\left. -v\left(\Pi_+X^{+\prime}+\Pi_-X^{- \prime}+
\pi_ff^{\prime}\right)\right] \ + \ boundary \ terms
                 \>.  \label{bff} 
\eea
An indirect way of checking the consistence of the above results can be 
carried out by evaluating the lagrangian density
on the space of solutions of the hamiltonian equations of motion.
It is easy to see from the free field form of the theory (\ref{bff})
 that it should be a total derivative.
 For models with a potential of the form $V(\phi)\propto\phi^a$, where $a$
 is an arbitrary real parameter, one can get the same results by using
 equations
 (\ref{bv}),(\ref{bvi}),(\ref{bviii}) to write (\ref{bi}) in the desired way.
 However for the Liouville model one has to explicitly solve the hamiltonian
 equations of motion to obtain the total derivative. Only for the models with
 $V\propto\phi^a$ it is possible to get the total derivative by direct 
 manipulation of the hamiltonian equations.\\

To finish this section we would like to analyze the problem of the equivalence 
of the quantum theory defined in the geometrical variables and the one which can be
defined in terms of the free fields. As is well-known in classical 
mechanics one can transform, via Hamilton-Jacobi theory, an arbitrary
 interacting theory into a trivial one.
 However, the canonical transformation which do that cannot be promoted into to
a unitary transformation relating the quantum wave functionals.
 This question has been analyzed in detail for the CGHS theory
\cite{Benedict} in the absence of matter fields. Moreover, in Ref. \cite{Louis2}
it was shown that the wave functionals obtained in terms of the CJZ variables
\footnote{The CJZ variables are only free field variables
  for the CGHS theory}
 for a generic 2D dilaton gravity \cite{Barvinsky}
 are equivalent to the wave functionals given in terms
  of the geometrical variables.
 Here we shall extend the analysis of \cite{Louis2}
 to show the equivalence of the quantum theory based on geometrical 
 variables and the quantization
obtained in the free field representation.
The basic idea to show the quantum equivalence is that in both representations
the constraints can be written in a form which is linear in momenta.
In the geometrical variables $(\rho,\phi)$, the constraints can be brought 
to the form 
 \cite{Louis}
 \be
 \pi_{\rho}=Q\left[C;\rho,\phi\right]\>,    \label{constraint1}
  \ee
  \be
  \pi_{\phi}={g\left[C;\rho,\phi\right]\over Q\left[C;\rho,\phi\right]}
  \>,             \label{constraint2}
  \ee
  where
  \be
  Q\left[C;\rho,\phi\right]=2\sqrt{\left(\phi^{\prime}\right)^2+
  \left[C-J\left(\phi\right)\right]e^{2\rho}} \>,
  \label{diii}
  \ee
  \be
  g\left[C;\rho,\phi\right]=4\phi^{\prime\prime}-4\phi^{\prime}
  \rho^{\prime}-2V(\phi)e^{2\rho}
  \>,  \label{function}
  \ee
  being the constant $C$ the ADM energy of the system.
  In the functional 
 Schr\"odin\-ger representation the quantum version of the above constraints
  have the solution
  \be
 \psi\left[ C;\rho,\phi\right] =\exp
 \left\{i\int dx\left[Q+\phi^{\prime}\log 
 \left({2\phi^{\prime}-Q\over 
 2\phi^{\prime}+Q}\right)\right]\right\}
 \>.        \label{wvi}
 \ee
On the other hand, the
 constraints in the free field variables become
 \be
 \pi_0=\pm r^1 \qquad  \pi_1=\pm r^0
 \>.  \label{constraint3}
 \ee
 We have to point now that, on the constrained surface, the 
 symplectic form of the theory turns out to be
 equal to the boundary term $\omega_b$ only.
 This is consistent with the fact that, in the absence of matter, 
 the theory is topological and it is described by a pair of canonically 
 conjugated global variables $(C,P)$. If we
 quantize the theory in the Schr\"odinger representation with wave functional
 $\Psi\left[r^a,P\right]$, the physical eigenstates of the hamiltonian operator are
 \be
 \Psi[r^a,P]=  \exp\left\{\pm{i\over2}
 \int dx \left[r^0\left(r^{1\prime}
 \right)-r^1\left(r^{0\prime}\right)\right]\right\}
 \exp iCP\>.       \label{wvii}
  \ee
The point is to show that both wave functionals (\ref{wvi}) and
(\ref{wvii}) are equivalent. To do that let us consider the generating functional
 $F\left[\rho,\phi,r^0,r^1,P\right]$ of the canonical transformation relating geometrical
and free field variables
\be
 \pi_{\rho}={\delta F\over \delta\rho}\qquad \pi_{\phi}={\delta 
 F\over \delta\phi}\>,\label{dxii}
 \ee
 \be
 \pi_{0}=-{\delta F\over\delta r^0}\qquad \pi_1=-{\delta F\over\delta
 r^1}
 \>.\label{dxiii}
 \ee
The wave functional $\psi\left[C;\rho,\phi\right]$ is determined by 
the equations
\be
 -i{\delta\over \delta\rho}\psi\left[C;\rho,\phi\right]
 =Q\left[C;\rho,\phi\right]
 \psi\left[C;\rho,\phi\right]
 \>.                          \label{dv}
  \ee
 \be
 -i {\delta\over\delta\phi}\psi\left[C;\rho,\phi\right]=
 {g\left[C;\rho,\phi\right]\over Q\left[C;\rho,\phi\right]}
 \psi\left[C;\rho,\phi\right]
 \>,     \label{dvi}
 \ee
\be
\hat M\psi=C\psi
\>,\label{dvii}
\ee
and $\Psi\left[C;r^a,P\right]$ by 
\be
 -i{\delta\Psi\over\delta r^0}=\pm r^1\Psi
\>,\label{setn}
\ee
\be
   -i{\delta\Psi\over\delta r^1}=\pm r^0\Psi
\>,     \label{och} 
  \ee
\be
\hat M\Psi=C\Psi
\>,\label{ochu}
\ee
We can use the constraint equations (\ref{constraint1}),(\ref{constraint2}) together with 
(\ref{dxii}) to find a relation $\rho=\rho(r^0,r^1)$
and $\phi = \phi(r^0, r^1)$ between the two sets of coordinates
\bea
{\delta F\over \delta \rho}&=&Q\left[C;\rho,\phi\right]
\>, \\
{\delta F\over \delta \phi}&=&{g\left[C;\rho,\phi\right]\over 
Q\left[C;\rho,\phi\right]}\>.\label{dxv}
\eea
 Because of the canonical transformation converts the constraints
 (\ref{constraint1}),(\ref{constraint2}) into (\ref{constraint3}) this relation is exactly the same that
 \be
 -{\delta F\over\delta r^0}=\pm r^1
 \>,
 \ee
 \be
 -{\delta F\over \delta r^1}=\pm r^0
 \>.
 \ee
  Now it is easy to see that the relation 
 between the quantum wave functionals
 is given for a generic model by
 \be
\Psi \left[ C;r^a,P \right] = \left[ \exp \left( -iF \left[ \rho,\phi,r^a,P \right] \right)
\psi \left[ C; \rho , \phi \right] \right] |_{\rho= \rho \left( r^0,r^1
\right) , \phi = \phi \left( r^0,r^1 \right) } 
  \>. \label{dxvi}
\ee 
It can be easily checked that $\Psi$ satisfies equations (\ref{setn}),(\ref{och}),(\ref{ochu})
whenever $\psi$ satisfies (\ref{dv}),(\ref{dvi}),(\ref{dvii}).
It is interesting to remark that the validity of the argument is based
on the fact that in both representations 
the constraints can be written in a 
form which is linear in momenta and therefore leads to
first order differential functional equations in the 
Schr\"odinger representation.

 \section{The CGHS model coupled to a Liouville Field}
One of the drawbacks of considering dilaton gravity models minimally
coupled to a scalar field in two-dimensions is that, unlike the realistic situation,
the matter field $f$ does not "feel" the gravitational field.
The purpose of the remaining part of this paper is to investigate theories
with a non-minimal coupling to the matter field.
If we consider the dimensional reduction of the action of a scalar field
minimally coupled to a 3D or 4D metric
\be
S_m\propto\int d^2x\sqrt{-g}\phi\left(\nabla f\right)^2
\>,\label{ci}
\ee
we see that matter couples to the dilaton, which plays the role
of the radial coordinate.
In this section we shall consider
an intermediate situation in which the 
matter couples to the metric
in a non-minimal way
but the coupling is independent of the dilaton field.
The simplest example is provided by the CGHS model 
coupled  to a Liouville field $f$
\be
S=\int d^2x\sqrt{-g}\left[R\phi+4\lambda^2-\beta\left(\nabla f\right)^2-\gamma e^{2\beta f}
-Rf\right]\>.\label{cii}
\ee
The pure gravity sector of this theory is canonically equivalent to two free fields and the
matter sector can also be mapped into a theory of two free fields.
We shall show that the full theory can be transformed into a 
theory of three free fields by combining
the metric-dilaton and the matter field in proper way.
This example provides a further evidence to the 
existence of a wide class of dilaton gravity
models coupled to a scalar field which are essentially equivalent to a theory of three free 
fields with a Minkowskian target space.

If we parametrize the metric as in (\ref{aiii}), 
the action becomes (\ref{bii}) in 
hamiltonian form where now the constraint functions are given by 
\bea
H&=&-{1\over2} \pi_{\rho}\pi_{\phi}+2\left(\phi^{\prime\prime}-\rho^{\prime}
\phi^{\prime}\right)-4\lambda^2e^{2\rho}+\beta\left(f^{\prime}\right)^2
\nonumber\\&&-
2\left(f^{\prime\prime}-f^{\prime}\rho^{\prime}\right)
+\gamma e^{2\left(\rho+\beta f\right)}+{1\over4\beta}\left(\pi_{\phi}
+\pi_f\right)^2
\>,\label{ciii}
\eea
\be
P=\pi_\phi \phi^{\prime}+\pi_{\rho}\rho^{\prime}-\pi_{\rho}^{\prime}
+\pi_f f^{\prime}
\>.\label{civ}
\ee
The equations of motion derived from the action (\ref{cii}) are
\be
\partial_+\partial_- \rho=0
\>,\label{cv}
\ee
\be
\partial_+\partial_-(\phi-f)=-\lambda^2e^{2\rho}+{1\over4}\gamma
e^{2\left(\rho+\beta f\right)}
\>,\label{cvi}
\ee
\be
\partial_+\partial_-\left(\rho+\beta f\right)=-{1\over4}
\gamma\beta e^{2\left(\rho+\beta f\right)}
\>,\label{cvii}
\ee
\be
\partial_{\pm}^2\phi-2\partial_{\pm}\phi\partial_{\pm}\rho=
\partial_{\pm}^2f-\beta\left(\partial_{\pm}f\right)^2
-2\partial_{\pm}f\partial_{\pm}\rho
\>.\label{cviii}
\ee
The general solution to the hamiltonian equations (\ref{cv}),(\ref{cvi}),(\ref{cvii}) 
can be
 expressed in terms of six arbitrary chiral functions
 (which are independent only before imposing the constraints)
  $A\left(x^+\right),
 a\left(x^+\right),P\left(x^+\right), B\left(x^-\right), b\left(x^-\right),$\newline 
 $Q\left(x^-\right)$
 as
 \be
 \rho={1\over2}\log\partial_+A\partial_-B
 \>,\label{cix}
 \ee
 \be
 \phi=-\lambda^2 AB+a+b
 \>,\label{cx}
\ee
\be
f={1\over2\beta}\log {\partial_+P\partial_-Q\over \left(1+{1\over4}\gamma\beta PQ\right)^2}-
{1\over2\beta}\log\partial_+A\partial_-B
\>.\label{cxi}
\ee
Guided by the solution (\ref{cix}),(\ref{cx}),(\ref{cxi}) we can write the
 following transformation to the new
 variables
 \be
 \phi=-\lambda^2AB+a+b
\>,\label{cxii}
 \ee
 \be
 \pi_{\phi}=-\left(\log A^{\prime}\right)^{\prime}+\left(\log B^{\prime}
 \right)^{\prime}
 \>,\label{cxiii}
 \ee
 \be
 \rho={1\over2}\log -A^{\prime}B^{\prime}
\>,\label{cxiv}
\ee
\bea
 \pi_{\rho}&=&2\lambda^2\left(A^{\prime}B-AB^{\prime}\right)-2a^{\prime}
 +2b^{\prime}+{1\over\beta}\left(\log P^{\prime}\right)^{\prime}
-{1\over\beta}\left(\log Q^{\prime}\right)^{\prime}
\nonumber\\&&
-{1\over\beta}\left(\log
A^{\prime}\right)^{\prime}+{1\over\beta}\left(\log B^{\prime}\right)^{\prime}
-{\gamma\over2}{P^{\prime}Q-PQ^{\prime}\over 1+{1\over4}\gamma\beta PQ}
\>,\label{cxv}
\eea
\be
f={1\over2\beta}\log {-P^{\prime}Q^{\prime}\over 
\left(1+{1\over4}\gamma\beta PQ\right)^2}
-{1\over2\beta}\log - A^{\prime}B^{\prime}
\>,\label{cxvi}
\ee
\be
\pi_f=\left(\log P^{\prime}
\right)^{\prime}-\left(\log Q^{\prime}\right)^{\prime}
-{\gamma\beta\over2}{P^{\prime}Q-PQ^{\prime}\over1+{1\over4}\gamma\beta PQ}
\>.\label{cxvii}
\ee
The two-form (\ref{bxiii}) can be written in terms of the new variables as
\bea
\omega&=&\int d x\left[-2\delta A\delta\left({a^{\prime}-m^{\prime}\over A^{\prime}}
\right)^{\prime}+2
\delta B\delta\left({b^{\prime}-n^{\prime}\over B^{\prime}}\right)^{\prime}+\right.
\nonumber\\&&\left.
+2\beta\left(\delta m+\delta n\right)\left(\delta m^{\prime}-\delta n^{\prime}
\right)\right]
\>,\label{cxviii}
\eea
where
\be
m={1\over2\beta}\left(\log P^{\prime}-\log A^{\prime}\right)
\>,\label{cxix}
\ee
\be
n={1\over2\beta}\left(\log Q^{\prime}-\log B^{\prime}\right)
\>,\label{cxx}
\ee
and the constraints $C_{\pm}$ are expressed 
\bea
C_{+}&=&-2a^{\prime}\left(\log A^{\prime}\right)^{\prime}
-{1\over2\beta}\left[\left(\log A^{\prime}\right)^{\prime}\right]^2
+2a^{\prime\prime}-{1\over\beta}\left(\log P^{\prime}\right)^
{\prime\prime}
\nonumber\\&&
+{1\over\beta}\left(\log A^{\prime}\right)^{\prime\prime}+
{1\over2\beta}\left[\left(\log P^{\prime}\right)^{\prime}\right]^2
\>,\label{cxxi}
\eea
\bea
C_-&=&2b^{\prime}\left(\log B^{\prime}\right)^{\prime}+{1\over2\beta}
\left[\left(\log B^{\prime}\right)^{\prime}\right]^2-2b^{\prime\prime}+
{1\over\beta}\left(\log Q^{\prime}\right)^{\prime\prime}\nonumber\\
&&
-{1\over\beta}\left(\log B^{\prime}\right)^{\prime\prime}
-{1\over2\beta}\left[\left(\log Q^{\prime}\right)^{\prime}\right]^2
\>.
\eea
If we now define
\be
X^+=-2A,\qquad \Pi_+=\left({a^{\prime}-m^{\prime}\over A^{\prime}}\right)^{\prime}
\>,\label{casa}
\ee
\be
X^-=2B,\qquad  \Pi_-=\left({b^{\prime}-n^{\prime}\over B^{\prime}}\right)^{\prime} 
\>,\label{coche}
\ee
\be
X^{f}=-\sqrt{2\beta}\left(m+n\right),\qquad \Pi_f=\sqrt{2\beta}\left(m^{\prime}
-n^{\prime}\right)
\>,\label{cxxiii}
\ee
the 2-form (\ref{cxviii}) becomes
\be
\omega=\int dx \left[\delta X^{+}\delta\Pi_++\delta X^-\delta\Pi_-
+\delta X^f\delta\Pi_f\right] \>,
\ee
indicating that the transformation
$\rho,\pi_{\rho},\phi,\pi_{\phi},f,\pi_f\longrightarrow
X^{\pm},\Pi_{\pm},X^f,\Pi_f$ is ca\-no\-nical.
It can be also checked that
\be
C_{\pm}=X^{\pm\prime}\Pi_{\pm}\pm{1\over4}\left(X^{f\prime}
\pm\Pi_f\right)^2 \>,
\ee
and therefore the canonical transformation converts the theory
into a theory of free fields.
Thus we have seen that, even when the coupling to the matter is not conformal,
it is possible to capture the canonical structure of the full theory in terms
of a set of free fields.
We must note that, although the free field variables $\left(X^f,\Pi_f\right)$ are
the standard ones of Liouville theory, the remaining pair of free fields $\left(X^{\pm},
\Pi_{\pm}\right)$ are not made out of purely dilaton gravity variables.
In the definition (\ref{casa}),(\ref{coche}) the fields $\Pi_{\pm}$ are functions of the
 Liouville field $f$ in addition to the $\left(\phi,\rho\right)$ fields.
 So the canonical transformation mixes intrinsically the three fields $\left(\phi,\rho;f\right)$  
to produce  three free fields.

\section{Free field representation of dilaton-coupled scalar models}

In this section we continue our analysis of the canonical structure 
of matter-coupled 2D dilaton gravity.
When a massless scalar field is coupled to gravity in higher
dimension and the theory is reduced to two dimensions by symmetry reduction
one derives an effective 2D dilaton gravity model with a dilaton-dependent
coupling.
The simplest model can be obtained from 3-dimensional gravity minimally
coupled to a massless Klein-Gordon field under the assumption of axi-symmetry
\be
ds^2_{(3)}=g_{\mu\nu}(t,r)dx^{\mu}dx^{\nu}+\phi^2(t,r)d\psi^2
\>,
\ee
with $x^0=t$ and $x^1=r$.
After dimensional reduction the matter-coupled 2D dilaton gravity theory
is described by the action
\be
S=\int d^2x\sqrt{-g}\left(R\phi-{1\over2}\phi\left(
\nabla f\right)^2\right)
\>,\label{ER}
\ee
where $f$ is the Klein-Gordon field. Remarkably the above model is also 
equivalent to the 4-dimensional
 Einstein-Rosen wave sector of general relativity. Imposing the
  cylindrical symmetry to the source-free 
Einstein theory
\be
ds^2_{(4)}=g_{\mu\nu}(t,r)dx^{\mu}dx^{\nu}+ \phi^2g_{ab}
(t,r)dx^adx^b \qquad  a,b=2,3
\>,
\ee
with $x^2=z$, $x^3=\phi$ and $det\ g_{ab}=1$, the reduced 
theory is a 2D dilaton gravity theory coupled to a 
$SL(2,R)/SO(2)$ coset space $\sigma$ model (see for instance \cite{korotkin}).
If the matrix $g_{ab}$ is assumed to be diagonal
 \be
 g_{ab}=\left(\begin{array}{cc}e^{-f}&0\\0&e^f\end{array}\right)
 \>,
 \ee
the solutions are the Einstein-Rosen gravitational waves 
and the 2D theory is given by (\ref{ER}).
If the 4D theory is general relativity minimally coupled to a scalar field $f$
 and we consider spherically symmetric fields like
(\ref{bii}) the spherical coordinates can be integrated out and,
after a conformal reparametrization of the 2D metric, the action becomes
\be
S=\int d^2x\sqrt{-g}\left(R\phi+{2\lambda^2\over\sqrt{\phi}}-{1\over2}
\phi\left(\nabla f\right)^2\right)
\>.\label{SS}
\ee
Note that this class of models are different from those considered in
\cite{korotkin} (and references therein), for which the matter sector is
given by a non-linear sigma model but the potential is trivial. By contrast
we consider models with a matter sector described by a single scalar field
but allows for a non-trivial potential. \\

An important ingredient in our analysis of minimally coupled models
 of section 3 is that the free field behaviour of the fields $\rho$
and $\phi$ of the theory (\ref{bi}) with a trivial potential can be extended to
 the models with an arbitrary
potential. Following the same argument we shall show in this section that the
free field representation of the Einstein-Rosen midisuperspace model, which we
shall explicitly construct in the next subsection, can be extended to a wide
family of models which includes the very important case of spherically
symmetric gravity (\ref{SS}).

\subsection{ Free field representation of the Einstein-Rosen midisuperspace}

In conformal gauge the equations of motion derived from the
action (\ref{ER}) are
\be
 \partial_+\partial_-\phi=0
 \>,\label{di}
  \ee
 \be
 4\partial_+\partial_-\rho+\partial_+f
 \partial_-f=0
\>,\label{dii}
\ee
\be
2\phi\partial_+\partial_- f+\partial_+\phi\partial_-f
+\partial_-\phi\partial_+f=0
\>,\label{laplace}
\ee
\be 
C_{\pm}=\pm 2\partial^2_{\pm}\phi\mp 4\partial_{\pm}\phi\partial_{\pm}\rho
\pm \phi\left(\partial_{\pm}f\right)^2=0
\>.\label{div}
\ee
Equation (\ref{di}) is a free field equation with solution
\be
\phi=A+B
\>,\label{FF}
\ee
where $A=A\left(x^+\right)$ and $B=B\left(x^-\right)$ are
arbitrary chiral functions.
Inserting this into (\ref{laplace}) we get
\be
2(A+B)\partial_+\partial_-f+\partial_+A\partial_-f
+\partial_-B\partial_+f=0
\>,
\ee
which can also be written
\be
2(A+B){\partial^2f\over \partial A\partial B}
+{\partial f\over\partial B}+{\partial f\over \partial A  }=0
\>,
\ee
or
\be
-{\partial^2 f\over\partial T^2}+{\partial^2 f\over\partial X^2}
+{1\over 
X}{\partial f\over \partial X}=0
\>,
\ee
where
\be
T={1\over2}(A-B),\qquad  X={1\over2}(A+B)
\>.   
\ee
This is a 2D Laplace equation in polar coordinates.
Its solution is known to be
\be
f={1\over2}\int_{-\infty}^{\infty} d\lambda\ 
J_0\left(
{\lambda\over2}(A+B)\right)
\left[A_+(\lambda)e^{i{\lambda\over2}(A-B)}+A_-(\lambda)
e^{-i{\lambda\over2}(A-B)}\right]
\>,
\ee
where $J_0$ is the zero order Bessel function and
$A_+^*(\lambda)=A_-(\lambda),\ A_+(-\lambda)=A_-(\lambda)$.
Finally we can use (\ref{dii}) to calculate $\rho$ as
\bea
\partial_+\rho &=& -{1\over4}\int_{-\infty}^{x^-}
\partial_+ f\partial_- f +\partial_+a
\>, \label{rhoi} \\
\partial_-\rho &=& {1\over4}\int_{x^+}^{+\infty}
\partial_+ f\partial_- f +\partial_- b
\>,\label{rhoii}
\eea
where we have introduced two new chiral functions
$a=a(x^+)$, $b=b(x^-)$.
The constraint equations can be written in the form
\be
C_+=2\partial_+^2A-4\partial_+A\partial_+a-2P
\>,
\ee
\be
C_-=-2\partial_-^2B+4\partial_-B\partial_-b+2Q
\>,
\ee
where the functions $P$ and $Q$ are given by
\be
P=-{\partial_+A\over2}\int_{-\infty}^{x^-}\partial_+ f
\partial_- f-{1\over2}(A+B)\left(\partial_+f\right)^2
\>,
\ee
\be
Q={\partial_-B\over2}\int_{x^+}^{+\infty}\partial_+ f\partial_- f
-{1\over2}(A+B)\left(\partial_- f\right)^2
\>.
\ee
Observe that in the gauge $A=x^+$, $B=-x^-$ and on the constraint surface
$C_{\pm}=0$ we recover the standard expression for the conformal factor
\be
\partial_{\pm} \rho=\pm {1\over4}\left(x^+-x^-\right)
\left(\partial_{\pm} f \right)^2
\>.
\ee

Due to Bianchi identities
$P$ must depend only on the coordinate $x^+$, so we can calculate its value
doing $x^-=-\infty$
\be
P=\lim_{x^-\rightarrow -\infty}-{1\over2}(A+B)
 \left(\partial_+f\right)^2
 \>,
 \ee
and an analogous argument leads to
\be
Q=\lim_{x^+\rightarrow \infty}-{1\over2}(A+B)\left(\partial_-f\right)^2
\>.
\ee
To explicitly calculate $P$, let us write
\bea
\partial_+f&=&{1\over4}\int_{-\infty}^{\infty}d\lambda\
\lambda\partial_+ A\left\{A_+(\lambda)
e^{i{\lambda\over2}(A-B)}\left[iJ_0\left({\lambda\over2}(A+B)\right.
\right.\right.\nonumber\\
&&\left.-J_1\left({\lambda\over2}(A+B)\right)\right]
-A_-(\lambda)e^{-i{\lambda\over2}(A-B)}\left[
 iJ_0\left({\lambda\over2}(A+B)\right)\right.\nonumber\\
 &&\left.\left.+J_1\left({\lambda\over2}(A+B)\right)\right]
\right\}
\>.\label{dp+}
\eea
In order to take the limit $x^-\rightarrow -\infty$ we 
shall assume that $B\left(x^-\right)$ is a monotonic
decreasing function which goes as $B\left(x^-\right)
\sim -x^-$ when $x^-\rightarrow -\infty$.
This requirement is necessary to preserve the meaning of $\phi$ as the radial coordinate.
If we substitute in (\ref{dp+}) the leading terms of the asymptotic expansions of the
Bessel functions when the argument goes to infinity
\be
J_0(x)\sim \left({2\over\pi x}\right)^{1\over2}\cos 
\left(x-{\pi\over4}\right)\qquad  |x|\rightarrow \infty
\>,
\ee
\be
J_1(x)\sim \left({2\over\pi x}\right)^{1\over2}\sin \left(x-
{\pi\over4}\right)
\qquad |x|\rightarrow  \infty
\>,
\ee
we get
\bea
\partial_+ f&\sim & {1\over2} \int_{-\infty}^{\infty}
d\lambda\left({\lambda\over\pi(A+B)}\right)^{{1\over2}}\partial_+A
\left[ A_+(\lambda)e^{i\lambda A}e^{-i{\pi\over4}}-\right.\nonumber\\&&
\left.A_-(\lambda) 
e^{-i\lambda A}e^{i{\pi\over4}}\right]
\>,
\eea
and therefore
\bea
P&=&-{1\over2}\left(
{i\over2} \int_{-\infty}^{\infty}
d\lambda\left({\lambda\over\pi}\right)^{{1\over2}}\partial_+A
\left[ A_+(\lambda)e^{i\lambda A}e^{-i{\pi\over4}}\right.\right.\nonumber\\&&
\left.\left. -A_-(\lambda) 
e^{-i\lambda A}e^{i{\pi\over4}}\right] \right)^2
\>.
\eea
In a similar way, and assuming that $A(x^+)$ is a monotonic increasing function
which goes as $A(x^+) \sim x^+$ when $x^+ \rightarrow +\infty$, it can be
proven that
\bea
Q&=&-{1\over2}\left( {i\over2}\int_{-\infty}^{\infty}d\lambda
\left({\lambda\over\pi}\right)^{{1\over2}}\partial_-B\left[ A_+(\lambda)e^{-i\lambda B}
e^{-i{\pi\over4}}\right.\right.\nonumber\\&&
\left.\left.-A_-(\lambda)e^{i\lambda B}e^{i{\pi\over4}}\right]
                              \right)^2
\>.
\eea
The above expressions for $P$ and $Q$ can be rewritten as follows
\be
P=-{1\over2}\left(\partial_+F\right)^2,\qquad  Q=-{1\over2}\left(\partial_-F\right)^2
\>,
\ee
where the field $F$ is given by 
\bea
F&=&{1\over2}\int_{-\infty}^{\infty}d\lambda \left(
\lambda\pi\right)^{-{1\over2}}\left[
A_+(\lambda)e^{i\lambda A}e^{-i{\pi\over4}}+A_-(\lambda) 
e^{-i\lambda A}e^{i{\pi\over4}}\right.
\nonumber\\&&\left.
+A_+(\lambda)e^{-i\lambda B}e^{-i{\pi\over4}}+
A_-e^{i\lambda B}e^{i{\pi\over4}}\right]
\>,\label{F}
\eea
and we must note that
the two chiral functions of the field $F=F_+\left(A\left(x^+\right)\right)
+F_-\left(-B\left(x^-\right)\right)$ are the same
\be
 F_+=F_-
 \>,\label{R}
 \ee
in other words, the field F is reflected at the boundary line $\phi = 0$. \\

For our problem it is convenient to define the symplectic form using the light-cone
$x^+=x^+_0$ and $x^-=x^-_0$ as the initial data surface
\be
\omega=\omega_++\omega_-=\int_{x^-=x^-_0} dx^+\delta j^-
+\int_{x^+=x^+_0} dx^-\delta j^+
\>,
\ee
where the light-cone components of the current $j^{\mu}$ can be easily calculated from the action
\be 
j^+=-4\phi\partial_-\delta\rho-\phi\partial_- f\delta f
\>,\label{jota+}
\ee
\be
j^-=4\partial_+\phi\delta\rho-\phi\partial_+f\delta f
\>.\label{jota-}
\ee
To evaluate $\omega$ explicitly we choose $x^+_0=\infty,\ x^-_0=-\infty$.
Taking into account 
(\ref{jota+}),(\ref{jota-}) and (\ref{FF}),(\ref{rhoi}),(\ref{rhoii})
 we arrive at
 \bea
 \omega&=&\omega_++\omega_-=\int_{x^-=-\infty} dx^+
-4\delta A\delta \partial_+a+\delta f\delta \left((A+B)
\partial_+f\right)\nonumber\\&&
+\int_{x^+=\infty}dx^- -4\delta B\delta\partial_- b+
\delta f \delta\left((A+B)\partial_-f\right)
\>.
\eea
Let us centrate ourselves in the calculation of $\omega_+$.
We shall work out first the differential of $f$
\bea
\delta f&=&{1\over2} \int_{-\infty}^{\infty}d\lambda -{\lambda\over2}
J_1\left({\lambda\over2}(A+B)\right)\delta A\left[A_+(\lambda)
e^{i{\lambda\over2}(A-B)}\right.\nonumber\\&&\left.
+A_-(\lambda)e^{-i{\lambda\over2}(A-B)}\right]
+J_0\left({\lambda\over2}(A+B)\right)\left[\delta A_+(\lambda)
e^{i{\lambda\over2}(A-B)}\right.
\nonumber\\&&
+A_+(\lambda)i{\lambda\over2}\delta A e^{i{\lambda\over2}(A-B)}+
\delta A_-(\lambda)e^{-i{\lambda\over2}(A-B)}\nonumber\\&&\left.
-A_-(\lambda)i{\lambda\over2}\delta A e^{-i{\lambda\over2}(A-B)}\right]
\>.
\eea
If we now substitute the asymptotic expansions of $J_0$ and $J_1$
we can factorize $\left(A+B\right)^{-{1\over2}}$ and the remaining terms can be 
collected to give
\bea
\delta f&\sim& {1\over2}\int_{-\infty}^{\infty} d\lambda 
\left(\pi\lambda(A+B)\right)^{-{1\over2}}
\delta \left[A_+(\lambda)e^{i\lambda A}e^{-i{\pi\over4}}+A_-(\lambda)e^{-i\lambda A}
e^{i{\pi\over4}}
\right]
\nonumber\\&&
= 
\left(A+B\right)^{-{1\over2}}\delta F_+
\>.
\eea
We need also to calculate
\bea
&\delta\left((A+B)\partial_+f\right)=&\nonumber\\
&={1\over2}\int_{-\infty}^{\infty} d\lambda {\lambda\over2}
\left\{\delta\left(\partial_+A A_+(\lambda)e^{i{\lambda\over2}(A-B)}\right)(A+B)\left[
iJ_0\left({\lambda\over2}(A+B)\right)\right.\right.&\nonumber\\&
\left. -iJ_1\left({\lambda\over2}(A+B)\right)\right]+\partial_+A A_+(\lambda)
e^{i{\lambda\over2}(A-B)}\left[iJ_0\left({\lambda\over2}(A+B)\right)\delta A\right.&\nonumber\\&
\left.-i{\lambda\over2}(A+B)J_1\left({\lambda\over2}(A+B)\right)\delta A-{\lambda\over2}
(A+B)J_0\left({\lambda\over2}\right)\right]&\nonumber\\&
-\delta\left(\partial_+ AA_-(\lambda)e^{-i{\lambda\over2}(A-B)}\right)(A+B)\left[iJ_0
\left({\lambda\over2}(A+B)\right)+J_1\left({\lambda \over2}(A+B)\right)\right]&\nonumber\\&
-\partial_+A A_-(\lambda)e^{-i{\lambda\over2}(A-B)}\left[iJ_0\left({\lambda\over2}(A+B)\right)\delta A
\right.&\nonumber\\&\left.\left.
-i{\lambda\over2}(A+B)J_1\left({\lambda\over2}(A+B)\right)\delta A
+{\lambda\over2}(A+B)J_0\left({\lambda\over2}(A+B)\right)\delta A
\right]\right\}\>,&
\eea
where we have used the identity
\be
J^{\prime}_1(x)=J_0(x)-{J_1(x)\over x}
\>.
\ee
Asymptotically
\bea
&\delta\left((A+B)\partial_+f\right)\sim&
\nonumber\\
&\sim{i\over2}\int_{-\infty}^{\infty} d\lambda \left({\lambda(A+B)\over\pi}
\right)^{{1\over2}}
\delta\left[\partial_+ A \left(A_+(\lambda)e^{i\lambda A}e^{-i{\pi\over4}}
+A_-(\lambda)e^{-i\lambda A}e^{i{\pi\over4}}\right)\right]&\nonumber\\
&\sim \left(A+B\right)^{{1\over2}}\delta \left(\partial_+ F\right)\>.&
\eea
In a similar way it can be checked
 that when $x^+\rightarrow \infty$
\be
\delta f\sim \left(A+B\right)^{-{1\over2}}\delta F_-
\>,
\ee
\be
\delta \left((A+B)\partial_- f\right)\sim \left(A+B\right)^{{1\over2}}\delta \partial_-F
\>.
\ee
With these results it is clear now that the symplectic form $\omega$ 
can be written in the following way
\bea
\omega&=&\int_{x^-=-\infty}dx^+ -4\delta A\delta 
\partial_+ a+\delta F_+\delta \partial_+ F
\nonumber\\&&
+\int_{x^+=\infty} dx^- -4\delta B\delta\partial_- b+\delta F_-\delta\partial_- F
\>.
\eea
In conclusion, defining
\be
X^+=A,\qquad \Pi_+=-4\partial_+a+2{\partial_+^2A\over\partial_+ A}
\>,
\ee
\be
X^-=B,\qquad   \Pi_-=-4\partial_-b+2{\partial_-^2B\over\partial_-B}
\>,
\ee
the constraints can be cast into the form
of a free field theory
\be
C_{\pm}=\pm\left(\partial_{\pm}X^{\pm}\Pi^{\pm}+
\left(\partial_{\pm}F\right)^2\right)
\>,
\ee
and the expression for $\omega$ 
\bea
\omega&=&\int_{x^-=-\infty}dx^+ \delta X^+
\delta \Pi_+ +\delta F_+\delta\partial_+ F\nonumber\\
&&+\int_{x^+=\infty}dx^- \delta X^-
\delta \Pi_-+\delta F_-\delta\partial_-F
\>,
\eea
shows that the transformation from the initial variables 
to the new ones 
$X^{\pm}, \Pi_{\pm}, F, \pi_F = \dot F$
is canonical.

We must note that the above free field representation has
 important consequences in the quantization of the theory.
 In contrast 
with pure dilaton gravity, for which the contributions to the
Virasoro central charge of the two free fields cancel in the 
Schr\"odinger quantization, we have now a completely different situation
and there is no way to avoid a non-vanishing center with $c=1$ coming from
the contribution of the field $F$
\be
\left[C_{\pm}(x),C_{\pm}(\tilde x)\right]=i\left(C_{\pm}(x)
+C_{\pm}(\tilde x)\right)\delta^{\prime}\left(x-\tilde x\right)\mp{i\over24\pi}
\delta^{\prime\prime\prime}\left(x-\tilde x\right)
\>, 
\ee
\be
\left[C_+(x),C_-(\tilde x)\right]=0
\>.
\ee
As a byproduct this explains why it has not been possible to find any
solution to the quantum constraints and the theory has been mainly
 studied in the framework of the reduced phase-space
\cite{Ashtekar} which sweeps the anomaly under the carpet.
Nevertheless, it is possible to modify the quantum constraints in such a way
 that the anomaly cancels.
The addition of a term depending on the fields $X^{\pm}$
\be
C_{\pm}(x)\pm {1\over48\pi}\left[{X^{\pm\prime\prime\prime}
\over X^{\pm\prime}}-\left({X^{\pm\prime\prime}\over
X^{\pm\prime}}\right)^2\right]
\>,\label{nueva}
\ee
produces a cancellation of the anomaly and allows to find states that are annihilated by all the
quantum constraints.
The above quantum modification of the constraints was introduced in 
Ref. \cite{Cangemi}
in the context of the CGHS theory and it was motivated by a mechanism based on an embedding-dependent 
factor ordering of the constraints \cite{Kuchar1}.
This type of quantum corrections were also introduced in Ref. \cite{Verlinde2}
to achieve a Virasoro algebra
 of central charge
$c=26$ also in the context of the (1-loop corrected) CGHS theory.
The modified constraints also
appeared in a more geometrical context.
The gravity part of these quantum constraints
\be
-iX^{\pm\prime}{\delta\over\delta X^{\pm}}\pm{1\over48\pi}
\left({X^{\pm\prime\prime\prime}\over X^{\pm\prime}}-\left(
{X^{\pm\prime\prime}\over X^{\pm\prime}}\right)^2\right)
\>,
\ee
are just the right-invariant vector fields of the Virasoro group with $c=-1$
\cite{Aldaya}, where the fields $X^{\pm},X^{\pm}\neq 0 $ are the diffeomorphism group parameters.

The modified quantum constraints
can be solved in terms of "gravitationally dressed" oscillators
defined by the relations
\be
\hat f\left(X\right)={1\over2\sqrt{\pi}}\int {dk\over |k|}e^{ikX}
\hat a\left(k\right)+\ c.c.
\>,
\ee
and it suggests to consider the gauge
 $X^{\pm}=\pm x^{\pm}$ which corresponds to the light-cone gauge of
critical string theory, as a natural one since the quantum modification of
(\ref{nueva}) vanishes. This might imply that the reduced phase-space
treatment of the theory \cite{Ashtekar},
which is based on the gauge-fixing $X^{\pm} = \pm x^{\pm}$,
could be equivalent to the proper Dirac quantization approach.
In the CGHS theory, the gauge $X^{\pm}=\pm x^{\pm}$
corresponds to the Kruskal gauge and the 1-loop reduced phase-space quantization
\cite{Mikovic2} in 
this gauge turns out to be compatible with the covariant approach.

\subsection{Models with non-vanishing potential}

Our purpose now is to study the theory obtained by adding a potential term
$\lambda^2 V\left(\phi\right)$ to the model (\ref{ER}). In the
hamiltonian formalism the constraints take the form
\bea
H & = & -\frac{1}{2} \pi_{\rho} \pi_{\phi} + 2(\phi'' - \phi' \rho') -
e^{2\rho} \lambda^2 V(\phi) + \frac{1}{2} \left( \frac{1}{\phi} \pi_{f}^2 +
\phi (f')^2 \right) \>, \\
P & = & \rho' \pi_{\rho} - \pi'_{\rho} + \phi' \pi_{\phi} + \pi_{f} f' \>.
\eea
We have seen that, for the theory with a vanishing potential, the central
charge in the Schr\"odinger representation is $c=1$, but we must stress that
this result can be maintained for an arbitrary potential. This follows
immediately from
the fact that the pure dilaton-gravity sector contributes with $c=0$ and the
remaining sector contributes with $c=1$, irrespective of the particular form
of the potential. So this suggests that the free field realization of the
Einstein-Rosen midisuperspace could be extended to models with a non-vanishing
potential. Using the transformation
which converts a generic pure dilaton gravity model into a free field theory,
the constraints become
\be
\partial_{\pm} \tilde X^{\pm}\tilde \Pi_{\pm}+{1\over2}\phi\left(\partial_{\pm}
 f\right)^2=0
 \>,
 \ee
although now the fields $\tilde X^{\pm},\tilde \Pi_{\pm}$ are not chiral
and $\phi$ in an involved function of
$\tilde{X}^{\pm},\tilde{\Pi}_{\pm}$.
Because of $\partial_{\mp}G_{\pm\pm}=0$, we can write
the constraints into the form
\be
\partial_{\pm}X^{\pm}\Pi_{\pm}+{1\over2}\left(\partial_{\pm}F\right)^2=0
\>,\label{ff}
\ee
if we define
\be
X^{\pm}=\tilde X^{\pm}|_{x^{\mp}=x^{\mp}_0},\qquad
\Pi_{\pm}=\tilde \Pi_{\pm}|_{x^{\mp}=x^{\mp}_0}
\>,
\ee
and
\be
\left(\partial_{\pm}F\right)^2=\phi\left(\partial_{\pm}
f\right)^2|_{x^{\mp}
=x^{\mp}_0}
\>,
\ee
where $x^{\pm}_0$ are arbitrary.
By analogy with the previous analysis of the Einstein-Rosen
midisuperspace the natural choice for $x_0^{\pm}$
are the surfaces at null infinity $x^{\pm}_0=\pm \infty$.
Assuming that there is not incoming (outgoing) matter
at null infinity $x^{+}=\infty$ $(x^-=-\infty)$
the fields $X^{\pm},\Pi_{\pm}$ coincide with the canonical free field variables of
 the pure dilaton gravity
 theory, as in the model with a vanishing potential.
Furthermore, the field $F$ is similar to the corresponding free field of the
Einstein-Rosen model (\ref{F}) and therefore inherits the reflecting property (\ref{R}).
It is important to stress now that a similar reflecting boundary condition was introduced 
in the one-loop corrected CGHS theory \cite{RST} due to the appearance of a time-like singularity curve.
The interaction of the matter field with the boundary line was a crucial ingredient in the approach of
Ref. \cite{Verlinde2} and now we have seen that the existence of a free field $F$ with
a reflecting boundary condition naturally emerges in the canonical analysis of the 
dilaton-coupled theory.

Up to now we have transformed the constraints 
into those of a free field theory where
the fields $X^{\pm},\Pi_{\pm}$ are chiral and 
the field $F$ satisfies the free field equation.
We want to prove now that this imply that the transformation going from 
the initial 
variables to the new ones $X^{\pm},\Pi_{\pm},F,\pi_F=\dot F$ is canonical.
To this end, let us consider the variation of the lagrangian density of the
theory under an infinitesimal 
diffeomorphism $\chi$
\be
\delta_{\chi} \left( \sqrt{-g}\cal L\right)=
\sqrt {-g}\nabla_{\mu}\left(\chi^{\mu}\cal L\right)
\>.
\ee
If the lagrangian of the theory is zero over solutions of the hamiltonian
equations of motion 
the above equation can be rewritten as
  \bea
&\delta_{\chi}\left(\sqrt{-g}\cal L\right)=&
\nonumber\\&
=-2G_{++}
\partial_-\chi^+-2G_{--}\partial_+\chi^-+
\partial_{\mu}j^{\mu}\left(\delta_{\chi}X^{\pm},\delta_{\chi}\Pi^{\pm}  
,\delta_{\chi} F\right)=0&
\>.\label{var}
\eea
This property can be easily checked when the potential is a power of 
the dilaton (as is the relevant case of spherically symmetric
Einstein gravity $V(\phi)\propto \phi^{-{1\over2}}$).
In this case the lagrangian
\be
{\cal L}  
  =
R\phi+\lambda^2\phi^a-{1\over2}\phi\left(\nabla f\right)^2
\>,
\ee
turns out to be a total derivative
 \be
{\cal L} = (1-a)\square \phi 
 \>,
 \ee
when restricted to the solutions of the hamiltonian equations
\be
R+\lambda^2a\phi^{a-1}-{1\over2}\left(\nabla f\right)^2=0
\>, 
 \ee
 \be
 \square \phi=\lambda^2\phi^a
 \>.
 \ee
 Therefore the equivalent lagrangian
 \be
 {\cal L}^{\prime}={\cal L} -(1-a)\square\phi
 \>,
 \ee
 satisfies the desired property and leads to the same canonical structure, up to boundary terms.
 If we choose the field $\chi$ in equation (\ref{var}) to be
  $\chi^+=x^-,\chi^-=0$, 
then we have
\be
2G_{++}=
\partial_{\mu}j^{\mu}=
\partial_+j^{+}
+\partial_-j^-
\>.\label{var1}
\ee
Now we assume that the constraints have the quadratic form (\ref{ff}).
So we arrive at
\be
-\Pi_+\partial_+X^+-\left(\partial_+ F\right)^2
=\partial_+j^++\partial_-j^-
\>,
\ee
Observing that
\be
\delta_{\chi} X^+=\chi^{\mu}\partial_{\mu}X^+=
x^-\partial_+X^+,\qquad   \delta_{\chi} X^-=0
\>,
\ee
\be
\delta_{\chi} \Pi^+=x^-\partial_+\Pi_+,\qquad    \delta_{\chi} \Pi_-=0
\>,
\ee
it is not difficult to see that the only possible expression for $j^-$ compatible with
(\ref{var1}) and which is a scalar density is
\be
j^-=-\Pi^+\delta X^+-\partial_+F\delta F_+
\>.
\ee
An analogous argument choosing $\chi^+=0, \chi^-=x^+$ shows that
\be
j^+=-\Pi^-\delta X^--\partial_-F\delta F_-
\>.
 \ee
Taking into account now that the Noether current $j^{\mu}$ can be interpreted
as the symplectic current potential for the 2-form $\omega$, where the
variation symbol $\delta$ plays the role of the exterior differential on
phase-space \cite{CPS}, we arrive at the conclusion that the symplectic form
\be
\omega=\int_{x^-=-\infty} dx^+\delta j^-
+\int_{x^+=\infty} dx^-\delta j^+
\>,
\ee
becomes
\bea
\omega&=&\int_{x^-=-\infty }dx^+\delta X^+
\delta \Pi_+ +\delta F_+\delta\partial_+ F\nonumber\\
&&+\int_{x^+=\infty}dx^- \delta X^-
\delta \Pi_-+\delta F_-\delta\partial_-F
\>,
\eea
and this shows that the free field variables $X^{\pm}, \Pi_{\pm}, F_{\pm},
\partial_{\pm} F$ are, in fact, canonical. Although the result only applies
when the potential is a power of the dilaton
we believe that a more detailed analysis could eliminate this restriction.

\section{Conclusions and Final Comments}
Stimulated by the paper \cite{CGHS} a lot of studies of 2D dilaton gravity and black holes have
been developed from different view points. In the absence of matter fields an exact 
Dirac quantization for all the dilaton gravity theories has been given in 
Ref. \cite{Louis}.
These theories admit a unified description in terms
 of Poisson-Sigma models \cite{Thomas}, which generalizes the gauge theoretical
formulation of the CGHS model \cite{CJ} and the well-known extra symmetry
of the later model can also be generalized
for a generic model \cite{CJNT}.
This provides additional reasons for the exact solvability of these theories
giving rise to a finite dimensional phase-space.
Moreover, the path-integral results are in accordance with this since the exact effective action
has been shown to exactly coincide with the classical one \cite{KKL}.
However, in the presence of matter fields, needed to have the Hawking radiation,
the theories are not longer topological and one must quantize an infinite number of degrees of freedom. For the
 CGHS model, and when the matter field
is minimally coupled to gravity, it is still possible to have control on the
 matter-coupled theory because the dilaton-gravity
sector of the theory can be canonically mapped into a set of two free fields with
signature (-1,1) \cite{Cangemi,Verlinde}.
Therefore, due to the fact that the matter sector itself is a free field the full 
theory is similar to a bosonic string theory
with a three dimensional Minkowski target space. In this
paper we have shown that this crucial property of the CGHS model remains
valid for a generic model of 2D dilaton gravity minimally coupled to a scalar field.
This way the well-known canonical equivalence of a Liouville field and a free field 
can be understood in a more general setting.
Furthermore the intriguing analogy of the Liouville field
as a longitudinal target space coordinate arising in no-critical string theory \cite{Polchinski}
can also be generalized for a generic theory of dilaton gravity. The longitudinal
target space coordinates $X^{\pm}$ are related to the logarithm of the conformal factor
and the dilaton.\\

The main drawback of considering theories with a minimal coupling to matter
is that they do not mimic properly the propagation of a scalar field on a
four-dimensionally geometry. In the general situation the matter
  field has a 2D dilaton coupling and, therefore, a natural question arises.
  Is still possible to map the matter-coupled theory into a free field theory?
  This question has been analyzed in this paper and the answer is in the affirmative. 
  For theories with a potential of the form $V(\phi)=\phi^a$, which includes
  spherically symmetric gravity, we have shown that a canonical transformation
converts the theory, up to a boundary term, into a free field theory with a Minkowskian
 target space.
This canonical equivalence emerges when one combine the previous result on
pure dilaton gravity and the explicit form of the free field representation of
the dilaton-coupled theory with a vanishing potential. The existence of a free
field $F$ which automatically incorporates the reflecting condition (\ref{R})
seem to indicate that could be an adequate variable to study the problem of
back reaction in the more realistic setting of dilaton-coupled models. We
shall explore this and related questions in future publications.

\section*{Acknowledgements}
J. C. acknowledges the Generalitat Valenciana for a FPI fellowship.
D. J. Navarro acknowledges the Ministerio de Educaci\'{o}n y Ciencia for a FPI fellowship.
We thank R. Jackiw and A. Mikovic for interesting suggestions.
We also would like to thank J. M. Izquierdo and M. Navarro for 
useful comments.

\end{document}